\documentclass[useAMS,usenatbib]{mn2e}
\usepackage{graphics, graphicx}

\newif\ifAMStwofonts
\def\deg{^{\circ}}

\def\eq{\begin{equation}}
\def\en{\end{equation}}

\def\lesssim{\raisebox{-0.3ex}{\mbox{$\stackrel{<}{_\sim} \,$}}}
\def\gtrsim{\raisebox{-0.3ex}{\mbox{$\stackrel{>}{_\sim} \,$}}}

\def\P3hat{{\mathaccent 94 P}_3}

\def\lapp{\ifmmode\stackrel{<}{_{\sim}}\else$\stackrel{<}{_{\sim}}$\fi}
\def\gapp{\ifmmode\stackrel{>}{_{\sim}}\else$\stackrel{>}{_{\sim}}$\fi}

\newcommand{\degrees}[1]{\ensuremath{#1^\circ}}

\title[Phase-locked modulation of PSR B1055--52]{Phase-locked modulation delay between the poles of pulsar B1055--52}
\author[Weltevrede, Wright \& Johnston]
{Patrick Weltevrede$^{1,2}$\thanks{E-mail: Patrick.Weltevrede@manchester.ac.uk}, Geoff  Wright$^{2,3}$ and Simon Johnston$^2$\\
$^1$Jodrell Bank Centre for Astrophysics, University of Manchester, Alan Turing Building, Manchester, M13 9PL, UK\\
$^2$Australia Telescope National Facility, CSIRO, P.O. Box 76, Epping, NSW 1710, Australia.\\
$^3$Astronomy Centre, University of Sussex, Falmer, Brighton, BN1 9QJ, UK}
\date{}
\pagerange{\pageref{firstpage}--\pageref{lastpage}}
\pubyear{2011}

\begin{document}

\maketitle
\label{firstpage}

\begin{abstract}
We present a detailed single pulse study of PSR B1055--52 based on
observations at the Parkes radio telescope. The radio emission is
found to have a complex modulation dominated by a periodicity of
$\sim{20}$ times its rotational period $P$ (0.197s), whose phase and
strength depends on pulse longitude. This periodicity exhibits a
phase-locked delay of about $2.5P$ between the main pulse (MP) and
interpulse (IP), presumed to be the opposite poles of the pulsar.
This delay corresponds to a light travel distance of many times the light cylinder radius.
More complex modulations are found within the MP on timescales down to
about $9P$, and both these and the principal modulation vary strongly
across the (at least) 7 components which the MP and IP exhibit. 
The nature of the single pulse emission, which ranges from smooth and longitudinally
extended to `spiky', is also component-dependent. Despite these
disparities, the total pulse intensity distributions at the MP and IP
are virtually identical in shape, suggesting a common emission
mechanism. In an attempt to account for the complex modulations we
examine a number of physical models, both intrinsic (which presuppose
the pulsar to be an isolated neutron star) and extrinsic (appealing to
the presence of circumstellar material to modulate the
emission). Significant objections can be made to each model, so this
pulsar's behaviour patterns remain a crucial challenge to theorists.
\end{abstract}

\begin{figure*} 
\includegraphics[height=0.99\hsize,angle=270]{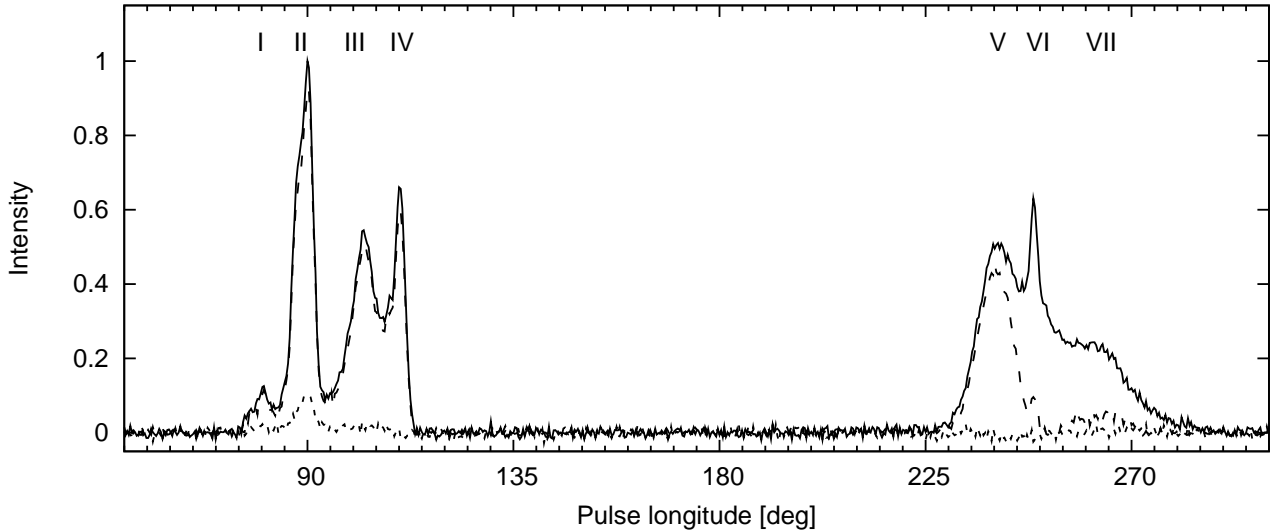}
\caption{\label{Fig1055profiles20}The pulse profile of PSR B1055--52
at a wavelength of 20~cm showing total intensity (solid line), linear
polarization (dashed line) and circular polarization (dotted line)
reproduced from Paper I. The component numbering convention used { in this paper} is
indicated.}
\end{figure*}

\section{Introduction}

PSR B1055--52 is a middle-aged pulsar (characteristic age about 0.5 Myr) known to emit pulsed emission from the radio to the $\gamma$-ray band. At radio frequencies it exhibits a complex integrated profile with separate main pulse (MP) and interpulse (IP) contributions, separated by roughly \degrees{180} in pulse longitude and argued to come from separate poles (\citealt{big90a,wqxl06}). In an earlier paper (\citealp[hereafter Paper I]{ww09}) the intensity pattern and polarization of the integrated profile were exploited to develop a self-consistent geometric model  within which the MP and IP emissions come from regions { associated with the opposing poles of the pulsar's magnetic axis, which is inclined to the rotation axis by about \degrees{75}.} 
PSR B1055--52 has recently been confirmed as a pulsed $\gamma$-ray source (\citealt{bt97,tbb99,aaa+10a}).  
The pulsed high energy components are then seen as emanating from higher altitudes in an outer gap.

{ An unexpected conclusion of Paper I was that, while the IP
emission comes from open field lines and its profile spans the
appropriate range of pulse longitudes, our oblique sightline towards
the pulsar requires that the more complex MP emission cannot originate
above a conventional polar cap. In fact, assuming the linear polarization can be interpreted with the rotating vector model \citep{rc69a} and a strict dipolar
field throughout the magnetosphere, the MP appears to be generated on
field lines normally associated with the pulsar's closed corotating
zone, suggesting that the true corotating zone is considerably smaller. Furthermore, the deduced location of the MP is asymmetrically placed within this zone.
The recent improved $\gamma$-ray
pulse profile (\citealt{aaa+10a}) spans the MP in its trailing half and an outer gap model is shown by these authors to be consistent with the geometry deduced from Paper I.
}

Our purpose here is to link our single-pulse radio data of PSR B1055--52 with the two-pole viewing geometry elucidated in Paper I. A pioneering study of this pulsar (at 645 MHz) was first published 20 years ago (\citealt{big90a}), identifying the principal { pulse profile} components and attempting to resolve the basic pulse modulations. { Evidence was found for periodic modulations, delayed correlation between the MP and the IP emission and for long-term changes in the profile at his observing frequency. Our intention here is to exploit the improved signal-to-noise ratio of our data to clarify and refine these results, albeit at a significantly higher observing frequency. We will demonstrate that each component has its own characteristic single-pulse behaviour, while exhibiting a subtle level of MP-IP interaction. As far as possible, we will attempt to indicate how these results, combined with the geometric constraints derived in Paper I, give insight into the physics of the pulsar system. }

For the reader's reference we reproduce the image of the integrated profile at 20 cm wavelength presented in Paper I (Fig. \ref{Fig1055profiles20}) where we were able to identify an additional component in the MP leading to a modified version of the component nomenclature introduced by Biggs.
 
Some years ago { an exercise similar to the present investigation} was undertaken with the pulsar B1702--19 (\citealt{wws07}), which, like PSR B1055--52, has relatively strong IP emission. In that pulsar the MP and IP were much narrower than those of PSR 1055--52 and the MP-IP separation was much closer to \degrees{180}. This was shown to result from the rotation of a highly inclined dipole field. { Single pulse data modulations of $\sim{11P}$ were found in} both the MP and IP emission which, crucially, exhibited an inter-pole time delay of about half a rotation period.  Communication between the polar regions is a fundamental clue to the underlying mechanism whereby the the pulsar emits. One may ask whether the communication is carried through the magnetosphere, the body of the neutron star or even via some external material. Our purpose here is to investigate whether PSR B1055--52 can help to answer some of these questions.
  
In the next section we present the record of our observations, followed by (in Section \ref{SectSingle}) a detailed account of the components' individual modulation. { In Section \ref{SectIssues} we discuss  the issues arising from our results and  consider alternative models.
Finally in Section \ref{SectSummary} we summarise our conclusions.}

\section{Observations}
\label{SectObs}

The observations of PSR B1055--52 presented in this paper were all
recorded at the Parkes radio telescope in Australia using different
backends and receiver systems. 
The main conclusions of this paper are based on a two hour long
observation recorded on 2007 April 23. All measurements presented in
this paper are based on this observation, unless stated otherwise. The
observation made use of the H-OH receiver centred at a
frequency of 1390 MHz with a bandwidth of 256 MHz. The signal was
processed on-line using the analogue filterbank, which combined the
two polarization channels of the linear feeds of the receiver to give
the total intensity (Stokes I) sampled once every 50$\mu$s for 512
separate 0.5 MHz wide frequency channels. The data were recorded on
tape for further off-line processing.

A second observation presented in this paper is a 1.9 minute long
observation which was recorded on 2008 July 25 during observations for
the Fermi timing program at Parkes \citep{wjm+10}. During this
observation the pulsar was brighter than usual, most likely because of
interstellar scintillation. The data were recorded in a very similar
way compared to the long observation, except that the data were
sampled at 80$\mu$s and that the centre beam of the multibeam receiver
was used at a centre frequency of 1369 MHz and a bandwidth of 256 MHz.

The third observation we made use of is an hour long observation
recorded on 2008 October 24. As for the short observation the
receiver used was the centre beam of the multibeam receiver at a
centre frequency of 1369 MHz and a bandwidth of 400 MHz. The data were
recorded using the BPSR backend \citep{kjv+10} with a sampling time of
64$\mu$s and a spectral resolution of 400 channels. The
observation was shortly interrupted to allow the feeds of the receiver
to be rotated by \degrees{90}. The two halves were appended off-line
to make a single one hour long observation. Unfortunately the pulsar
was relatively weak during this observation.

The first step in the off-line data reduction involved the masking of frequency channels affected by radio frequency
interference (RFI). The remaining frequency channels were added
together accounting for the dispersion delay caused by the interstellar
medium using the {\sl SIGPROC} software
package\footnote{http://sigproc.sourceforge.net/}. Using the known
pulse period the resulting dedispersed time series were folded
following the procedure outlined in \cite{wes06}. The data are now
arranged in a { two} dimensional array (pulse-stack) where the rotational
phase (pulse longitude) of the star is one axis and the pulse number
the other.

Although narrow band RFI was removed from the data, there was still
some impulsive broad-band RFI affecting the 2 hour observation. The
data were therefore inspected by eye and 2\% of the pulses in the
pulse-stack were flagged and ignored during further analysis.

The next step in the off-line data reduction attempts to correct what
we will refer to as ``baseline artifacts''. These artifacts arise when
relatively strong signals are recorded using the analogue filterbank
which records the data with 1-bit sampling. The result is that the
off-pulse intensity (baseline) is not constant throughout the
observation, but is decreased both before and after a strong signal is
recorded\footnote{
The reason why the noise level before a strong signal in a frequency
channel can be affected is the combination of the earlier arrival of
the strong signal at frequency channels corresponding to higher sky
frequencies in combination with frequency channels sharing
filters.} The baseline artifact is not severe in the two hour long
observation,
but it is much stronger in the
1.9 minute observation during which the pulsar appeared significantly brighter and for which the sampling rate was slightly lower.

As a result of the baseline artifact adjacent pulse longitude
bins are not entirely independent of each other, which is
undesirable.  { The baseline artifact} can be partially removed\footnote{ It is impossible to remove the baseline artifact
perfectly, because of its complex time-frequency-intensity
behaviour.} by convolving the
recorded (dedispersed) signal with two exponential functions, one removing the
decrease in the noise level after a strong signal and one removing the
noise level before a strong signal. After this procedure the off-pulse
baseline no longer showed the characteristic dips directly before and
after the pulse.
Finally a constant
baseline was subtracted from each pulse to make the average intensity
in the off-pulse region zero.

Although we do not expect the adjacent pulse longitude bins to be
completely independent of each other after the correction, at least it
gives us a tool to test if the baseline artifacts can be expected to
influence our results. We find no appreciable differences in any of
the results presented in this paper when using the corrected or
uncorrected data. This suggests that the correlations that will be
discussed are intrinsic to the pulsar rather than instrumental. In
addition the BPSR data, which did not suffer from baseline artifacts,
are fully consistent with the results presented in this paper.

\section{Single pulse modulation}
\label{SectSingle}

\subsection{Pulse-stack}

\begin{figure}
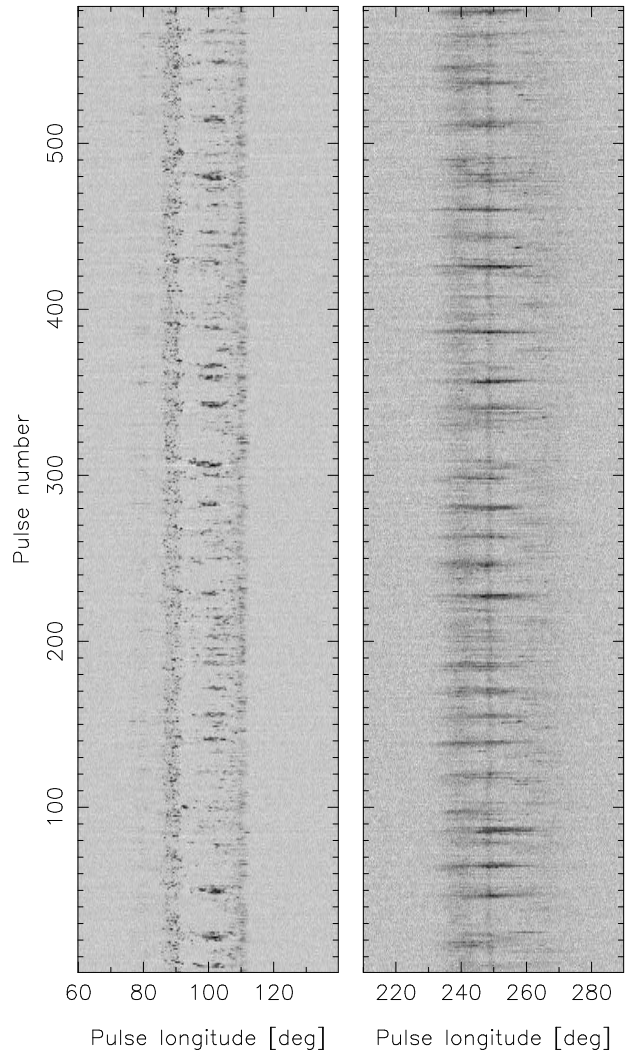

\includegraphics[width=1.65\hsize,angle=270]{1055MPstackNew.ps}
\hspace*{0.01\hsize}
\includegraphics[width=1.65\hsize,angle=270]{1055IPstackNew.ps}
\caption{\label{FigSinglePulses}{\em Left:} A pulse-stack of 581
successive pulses of the MP observed when the pulsar was exceptionally
bright. {\em Right:} The pulse-stack of the same pulses as in the left
panel, but now centred at the IP. The $\sim20P$ modulation pattern is
visible in the central components of the MP and across the IP, as is
the delay of $\sim 2.5P$ by which the pattern observed in the MP lags
the IP.}
\end{figure}

Before embarking on a detailed analysis of the statistical properties
of the emission of PSR B1055--52, it is good to first get a basic
understanding of its behaviour by inspecting some of its individual
pulses by eye. The pulse-stacks of the MP and IP of PSR B1055--52 are
plotted in Fig. \ref{FigSinglePulses} with the pulse longitude scale
aligned with that of Fig. \ref{Fig1055profiles20} by placing the
profile peak at pulse longitude \degrees{90}. As one can see, the
pulse intensities are highly variable from pulse to pulse. The most
obvious modulation pattern is the roughly periodic brightening which
happens on a timescale of $\sim20P$. This modulation is especially
visible in the IP and the third component of the MP (at
$\sim\degrees{100}$ pulse longitude).

Periodicities in pulse intensity are a common phenomenon in radio
pulsars (e.g. \citealp{wes06,wse07}), often associated with drifting
subpulses \citep{dc68}, although in the case of PSR B1055--52 the periodicity 
arises from intensity modulations rather than phase modulations. We
will show in Sect. \ref{SectPhaseCorr} that there exists a stable phase
delay between the modulation patterns observed in the MP and IP, a
feature which in fact is detectable by eye in
Fig. \ref{FigSinglePulses}.

The various components of the pulse profile have completely different
characteristics in their single pulse behaviour. It is particularly
striking that the pulses of the MP have a very different appearance to
those of the IP. The pulses of the MP show much sharper
structures and can have widths much smaller than the profile component
widths. In contrast the individual pulses of the IP are much more
extended or smeared out in pulse longitude. The pulses of the MP, especially those of component
II (pulse longitude $\sim\degrees{90}$), can be characterised as very
spiky, i.e. the pulse-stack at the location of component II is dotted
with narrow spikes of emission with high peak fluxes thereby resembling the
emission observed, for instance, in PSR B0656+14
\citep{wws+06}. Furthermore, it appears that not every MP component
shares the $20P$ modulation as clearly as some of the others.

Also within the IP the individual components have differing
single-pulse characteristics.  Although component VI (pulse longitude
$\sim\degrees{248}$) participates in the modulation cycle, it never fades
as much as the other IP components, thereby forming a vertical
band in the pulse-stack. In addition, component V (pulse longitude $\sim\degrees{240}$) and VII (pulse longitude $\sim\degrees{260}$) while also subject to modulation, do
not completely disappear.
The middle component (VI) appears to form some sort of boundary in the
sense that sometimes the leading side of the IP profile participates
much more strongly in the modulation cycle compared to the trailing
side (e.g. around pulse 96 and 245) and sometimes this appears to be
reversed (e.g. around pulse 85 and 355). Nevertheless in many cases
both components clearly show the same cycle. The duty cycle of the
$\sim20P$ modulation pattern is relatively small in both the MP and IP
with the pattern being bright for only a few pulses.

\subsection{\label{SectFlucSpectra}Fluctuation spectra}

\begin{figure}
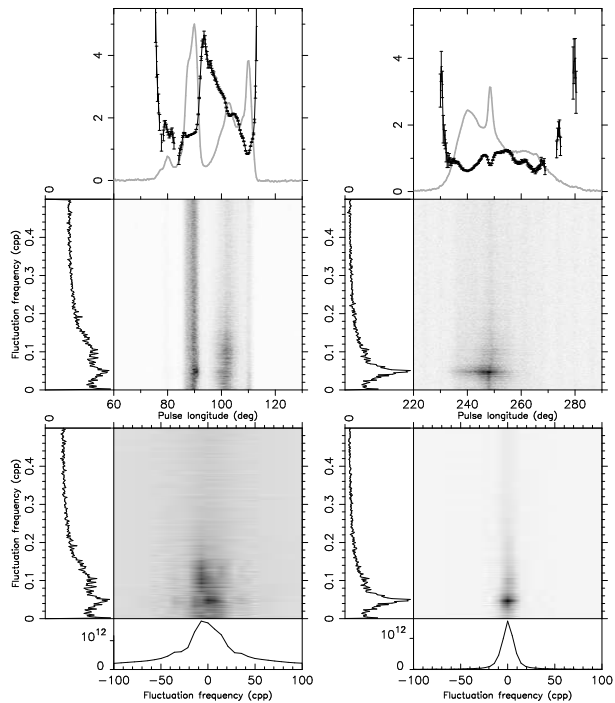

\begin{center}
\includegraphics[width=1.1\hsize,angle=270]{SPW001_0101SpectraMP.ps}
\includegraphics[width=1.1\hsize,angle=270]{SPW001_0101SpectraIP.ps}
\end{center}
\caption{\label{FigModulation}Modulation index and fluctuation spectra
of PSR B1055--52. The left-hand plots are for the MP and the
right-hand plots for the IP. {\em Upper panel:} The peak of the pulse profile (grey
curve) is normalised to 5 and the longitude resolved modulation index
is shown as the points with error bars. {\em Middle panel:} The
LRFS. {\em Lower panel:} The 2DFS. The data is re-binned to 1000
pulse longitude bins.}
\end{figure}

The time-averaged properties of the periodic subpulse modulation can
be studied in more detail by calculating fluctuation spectra. We will
follow the methods developed by \cite{es02} and for more details about
the analysis we also refer to \cite{wes06}. The first
type of fluctuation spectrum computed is the longitude-resolved
fluctuation spectrum (LRFS; \citealt{bac70b}). This fluctuation
spectrum is obtained by separating the pulse-stack into blocks of 256
pulses long. By calculating Fourier transforms along constant pulse
longitude columns in the pulse-stack one gets information about the
periodicities present for each pulse longitude bin. By averaging the
{ square of the} spectra obtained for the consecutive blocks of data the LRFS is
obtained. These spectra are shown in the middle panels of Fig.
\ref{FigModulation}.

The LRFS confirms and clarifies the features noted by eye in
the pulse-stack. The LRFS of the IP clearly shows the $\sim20P$
periodicity, which corresponds to $\sim0.05$ cycles per period
(cpp). The same periodicity is visible in the LRFS of the MP
(especially in the trailing half of component II at \degrees{90} and
component III at \degrees{100}), although there is much more
additional modulation power at different frequencies (or timescales)
compared to the IP. This is consistent with the $\sim20P$ { periodicity} visible most
clearly in the pulse-stack of the IP. As can be seen in the side panel
of the LRFS of the MP, the spectral power appears to be a combination of a
white-noise like plateau combined with a peak corresponding to the
\mbox{$\sim20P$} periodicity and a red-noise like spectrum corresponding to
timescales larger than $\sim5P$ (corresponding to frequencies smaller than
$\sim0.02$ cpp). The different profile components show different mixtures of
these spectral behaviours.

The two dimensional fluctuation spectrum (2DFS; \citealt{es02}) is a
different type of fluctuation spectrum that can reveal additional
information about periodic subpulse modulation. The vertical frequency
axis is very similar to the case of the LRFS, which can be
expressed as $P/P_3$, where $P_3$ corresponds to the $20P$ periodicity
visible in the pulse-stack. The horizontal axis
of the 2DFS denotes a frequency as well, which can be expressed as
$P/P_2$, where $P_2$ corresponds to the horizontal separation of the
subpulses. If the power were concentrated at positive values of
$P_2$ it would indicate that subpulses tend to shift (or drift) to later pulse
longitudes in successive pulses.

The IP shows the same $P_3=21\pm2P$ periodicity (for a
discussion about error bars on fluctuation frequencies, see
\citealt{wes06}) and its 2DFS is perfectly symmetric about the
vertical axis. This indicates that subpulses in successive pulses do
not drift on average to later or earlier pulse longitudes, which is consistent
with the fact that the intensity bands in the pulse-stack of the IP
appears to be horizontal. The 2DFS of the MP has a much more complex
shape. As expected from inspection of the LRFS, 
the 2DFS of
the MP also  shows power corresponding to a wider range of periodicities. There is horizontal structure visible
in the 2DFS with power concentrated on frequencies corresponding to
the component separations within the MP. The asymmetry of the 2DFS
about the vertical axis indicates that the appearance of subpulses in
the different components is correlated, although the correlation is
complex in nature. The possible pulse longitude drift of subpulses
participating in the $P_3\simeq20P$ cycle will be explored in more
detail in Section \ref{SectPhaseCorr}.

\subsection{Stability of the $P_3\simeq20P$ cycle}
\label{SectStability}

\begin{figure}
\includegraphics[height=0.95\hsize,angle=270]{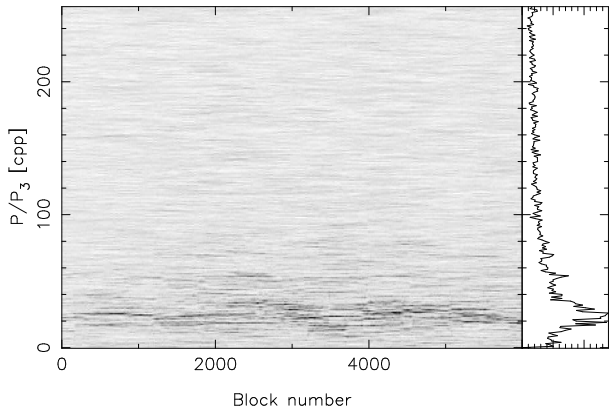}
\caption{\label{FigS2DFS}A section of the S2DFS showing the evolution
of $P_3$ by sliding a 256-pulse long window over 6255 successive
IP pulses, thereby generating 6000 spectra.}
\end{figure}

The peaks in the LRFS and 2DFS caused by the $P_3\simeq20P$ cycle is
broad, hence the periodicity does not appear to be particularly
stable. Although the first signs are not promising, we pursued the
possibility that a very precise and stable clock is responsible for
the observed periodicity, although modulated by a noise-like
process. The presence of such a stable signal would have important
consequences, for instance when the periodicity is linked to a
possible fallback debris disks \citep{cs08}.

The possible presence of a stable periodicity among other time-varying
signals can be tested by making use of the sliding 2DFS (S2DFS;
\citealt{ssw09}). The S2DFS is a time-resolved version of the 2DFS and
is therefore ideal to analyse the time-dependence of the
$P_3\simeq20P$ cycle. The S2DFS is calculated by computing the 2DFS
for blocks of 256 pulses with the start of the block shifting with one
pulse at the time. In other words, the first 2DFS is computed from
pulses $1-256$, the second for pulses $2-257$, etc. The two
dimensional spectra are collapsed over the $P_2$ axis, giving rise to
a one dimensional spectrum for each block. By stacking the resulting
spectra for the successive blocks of data one obtains the S2DFS (see
Fig. \ref{FigS2DFS}). The spectra obtained for these ``block
numbers'' are not independent of each other, as the time resolution is
determined by the size of the sliding window of 256 pulses. The S2DFS
shows a constantly changing frequency of the modulation pattern and no
fixed periodicity is present among the other time-variable
fluctuations. The second conclusion is that the timescale of the
time-variability of the modulation pattern is short, i.e. a few
hundred pulses or less.

Secondly we tested the hypothesis that the observed $P_3\simeq20P$
cycle is a very stable periodicity affected by a large phase-noise.
In other words, could it be possible that the bright intensity bands
in the pulse-stack appear often a few pulses earlier or later then
what would be predicted by an otherwise strictly periodic cycle? In some ways this
is analogous to what is assumed for pulsar timing experiments, where
the pulse arrival times are compared to a timing model containing well defined periodicities, giving rise to
timing residuals with relatively small amplitude. By making use of the
timing software developed by \cite{wje11} we tried to model the
measured pulse numbers for which we observe a bright intensity band in
the IP with a timing model being a fixed frequency. This frequency was
initially set to the frequency of the peak in the LRFS obtained for the same
stretch of data of a few hundred pulses long which was used to measure
the pulse numbers corresponding to the appearance of the intensity bands. It turns out to be
impossible to get phase connection of the measured arrival times of
the intensity bands, even after changing the frequency in the timing
model. This confirms the suspicion that the periodicity of the
modulation pattern is unstable, rather being a fixed value modulated
by phase noise with an amplitude less than $20P$.

\subsection{\label{SectModIndex}Modulation index}

The upper panels of Fig. \ref{FigModulation} show, besides the pulse
profile (grey curve), the longitude resolved modulation index of the
MP and IP. The modulation index is a normalised measure of the
variability of the pulse intensities at a given pulse longitude and is
computed from the LRFS (see e.g. \citealt{es03,wes06}). The error bars
on the modulation index was derived by bootstrapping the data
(pulse-stack). For each iteration in the bootstrap method an equivalent
pulse-stack was generated by adding random noise samples with an
root-mean-square (rms) equal to that measured in the off-pulse
region. The standard deviation of the distribution of modulation
indices obtained in this way is set to be the (1-$\sigma$)
error bar. Simulations with artificial pulse-stacks indicate that the error bars derived in this
way are more reliable than the analytically estimated error bars as
computed for instance by \cite{es03} and \cite{wes06}.

Typically one measures modulation indices of around $0.5$ (at
observing frequencies around 1400 MHz; e.g. \citealt{wes06}) at the centre
of components with the modulation index flaring up at the component
edges. A similar flaring up at the profile edges is seen for this
pulsar. The most variable pulse intensities (i.e. highest modulation
index) are seen between components II and III of the MP.  The value
of the modulation index in this region of the MP is as large
as that observed for PSR B0656+14 at a frequency of 327 MHz
\citep{wws+06}. However one cannot direct compare these values because
the high modulation index of PSR B1055--52 is observed between profile
components, while that of PSR B0656+14 is observed at the profile
centre (which is what makes the high modulation index of PSR B0656+14
unusual). The modulation index derived from the BPSR data is similar,
suggesting that the high modulation index is not caused by the
baseline artifact.

The modulation index profile of the IP is on average lower than that
of the MP. This is not surprising given the much more spiky appearance
of the pulse-stack of the MP.  The modulation index profile shows
a dip at the location of component VI, which is caused by the
presence of steady emission in that component (as is seen as the
vertical band in the pulse-stack in Fig. \ref{FigSinglePulses}).

\subsection{\label{SectVariability}Profile variability}

\begin{figure}
\begin{center}
\includegraphics[height=0.99\hsize,angle=270]{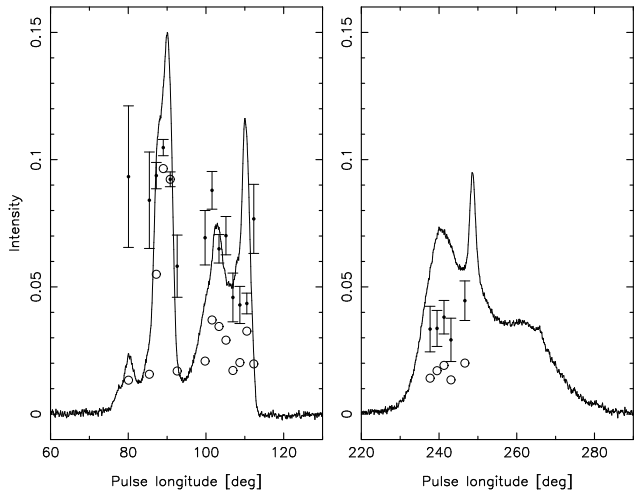}
\end{center}
\caption{\label{FigVariability}{ The pulse profile variability
of $2000P$ long sub integrations of the MP {\em (left)} and IP {\em
(right)} is quantified by the longitude resolved standard deviation (open circles) and the modulation index (points with error bars). The peak of the average pulse profile (solid line) is normalised to a value of 0.15.}}
\end{figure}

\cite{big90a} reports that the pulse profile of PSR B1055--52 is
variable at relatively long timescales in the sense that the peak
amplitudes of the different components fluctuate when integrations of
$2000P$ are taken. { In order to quantify this effect we averaged pulses in blocks of 2000 pulses. Using the methods outlined in the previous subsection we calculated the longitude resolved modulation index and standard deviation (after rebinning the data to 200 pulse longitude bins). As can be seen in Fig. \ref{FigVariability}, the highest variability (the open circles indicating the standard deviation of the intensities at a given pulse longitude) is observed in component II (pulse longitude $\sim\degrees{90}$). However, after normalisation with respect to the average intensity observed at these longitudes (i.e. the modulation index; the points with error bars) one can see this variability is not exceptional compared to other components in the MP. The emission of the IP is more stable resulting in a low modulation index for the profile variability which is comparable to that observed in component IV in the MP (pulse longiture $\sim\degrees{110}$). This is consistent with the more spiky nature of the emission observed for the MP, hence longer stabilisation timescales can be expected as is observed for PSR B0656+14 \citep{wws+06}. }

Unlike \cite{big90a}, we find that component II has the largest
amplitude throughout the observation, while \cite{big90a} observed
component III to have in general the largest amplitude\footnote{In the
component name convention of \cite{big90a} this corresponds to
component I and II rather than components II and III.}.  We have found
no exception to this rule in 67 observations of this pulsar between
2007 and 2011 for the Fermi timing program at Parkes
\citep{wjm+10}. It must be noted that our observing wavelength (around
$20$ cm) is different from that in \cite{big90a} (around $50$ cm), but
the profile evolution between these frequencies is not very strong
(see e.g. Fig. 2 in \citealt{ww09}). We attribute these differences
in the profile to instrumental effects in the observations of
\cite{big90a}, which had a much lower time resolution. Later Parkes radio
observations as shown by for instance \cite{wml91,vdhm97,tbb+99} are consistent with the observations presented in this paper.

\subsection{Pulse-energy distribution}

\begin{figure*}
\includegraphics[height=0.95\hsize,angle=270]{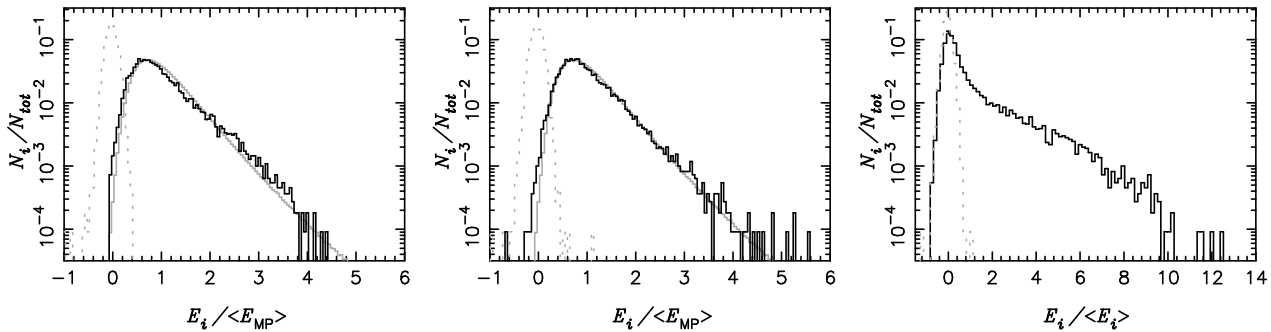}
\caption{\label{FigEnDist}The pulse energy distribution of the MP {\em
(left panel)} and the IP {\em (central panel)}. The on-pulse energy
distribution is plotted as the black histogram and the off-pulse
distribution as the grey dashed histogram. The pulse energies are normalised
to the average MP pulse energy. The lognormal fit for the IP (solid grey histogram) is shown in both the left and central panel. The pulse energy distribution of component III is
plotted in the {\em right panel}, where the pulse energies are
normalised to the average pulse energy in that component.}
\end{figure*}

The fluctuation spectra have { shown} us in some detail about
periodicities in the pulse intensity. The modulation index tells us,
in a very limited way, how variable the pulse intensities are. More
detailed information about the intensity variations can be obtained by
making a pulse energy distribution. The pulse energy distribution is
calculated by integrating the intensities as recorded in the pulse-stack over a pulse longitude range. This has been done for
the full pulse longitude range of the MP and IP separately, which
gives a pulse energy $E_i$ for each stellar rotation $i$. Given that
the observations were not flux calibrated, only the values $E_i$
relative to each other are meaningful. We therefore choose to
normalise the pulse energies $E_i$ by $\left<E_\mathrm{MP}\right>$, the average pulse energy
measured for the MP. Closer inspection of the data reveals that
the flux of PSR B1055--52 is not constant throughout the observation
because of interstellar scintillation. The pulsar is
weak at the start and end of the observation, but has a nearly constant flux
for 35 minutes. The energy distributions shown in Fig.
\ref{FigEnDist} are therefore based on these 35 minutes, which
much more accurately reflect the intrinsic pulse energy
distribution of the source. The off-pulse energy distributions (noise
distributions) are shown as well in order to quantify the uncertainty
in the pulse energy measurements (dashed grey distributions). The off-pulse energy distribution is
calculated by integrating the off-pulse region over a pulse
longitude window with equal width compared to the on-pulse
window.

Using the methods explained in \cite{wws+06} we have fit the shape of
the energy distribution of the IP by a log-normal distribution
convolved with the observed noise distribution. The parameters
of the lognormal distribution
\begin{equation}
P(E) = \frac{\left<E\right>}{\sqrt{2\pi}\sigma E}\exp\left[-\left(\ln\frac{E}{\left<E\right>}-\mu\right)^2/\left(2\sigma^2\right)\right]
\end{equation}
we fit for are found to be $\mu=-0.13$ and $\sigma=0.51$. This fit matches
the observed energy distribution of the IP very well (see centre panel
of Fig. \ref{FigEnDist}).

The fit for the IP energy distribution is reproduced in the left panel
of Fig. \ref{FigEnDist} (without any scaling). As one can see the
shapes of the energy distribution observed for the MP and IP (centre
panel of Fig. \ref{FigEnDist}) are very similar.  This close
resemblance is surprising given the completely different
characteristics of the pulses as can be seen in the pulse-stack (see
Fig. \ref{FigSinglePulses}) and given the conclusion in Section
\ref{SectFlucSpectra} that the fluctuation spectra of especially the
MP can be described by a mixture of different emission types. { The
resemblance of the MP and IP energy distribution is confirmed in
other (shorter) observations.}

There are small differences in the shape of the energy distribution of the MP and IP.
The IP shows a tail beyond $E_i\,\gtrsim
4\left<E_\mathrm{MP}\right>$ which is not present in the MP and a
slight excess of MP pulses with energies in the range
$2\left<E_\mathrm{MP}\right>\,\lesssim E_i\,\lesssim
4\left<E_\mathrm{MP}\right>$. In addition the energy distribution of
the MP is skewed to slightly lower energies at the low-energy part
compared to that of the IP.
Closer inspection reveals that the energy distributions of most of the individual components are very
similar to the shape of the overall energy distribution of the MP and IP.
The exception is component III, which shows a clear
break in its pulse energy distribution (see right panel of Fig.
\ref{FigEnDist}) and component I, which is too weak to accurately
determine its pulse energy distribution. 
The different shape of the pulse energy distribution of component III is therefore responsible for the slight difference in the 
overall energy distribution of the MP and IP. 

{ Motivated by the similarity between the pulse energy distribution of the MP and IP, we investigated if the pulse energies of the MP and IP are correlated at timescales  slightly longer than the modulation cycle. Following a similar procedure employed by \cite{wws07} for PSR \mbox{B1702--19}, we started by averaging blocks of 20 pulses. This ensured that the modulation pattern is no longer present in the signal, which was confirmed by calculating the LRFS. MP and IP energies were calculated for each block by integrating over the full pulse longitude range over which emission was detectable in the pulse profile. Finally, by cross correlating the MP and IP sequence of energies it was possible to investigate if the power output of the MP and IP are related. However, there was very little evidence for correlation between the MP and IP power output on timescales larger than the modulation cycle, apart from long-term correlations caused by the fact that both the MP and IP are affected simultaneously by interstellar scintillation. }

\subsection{\label{SectPhaseCorr}Phase-locked modulation}

\begin{figure*}
\includegraphics[height=0.95\hsize,angle=270]{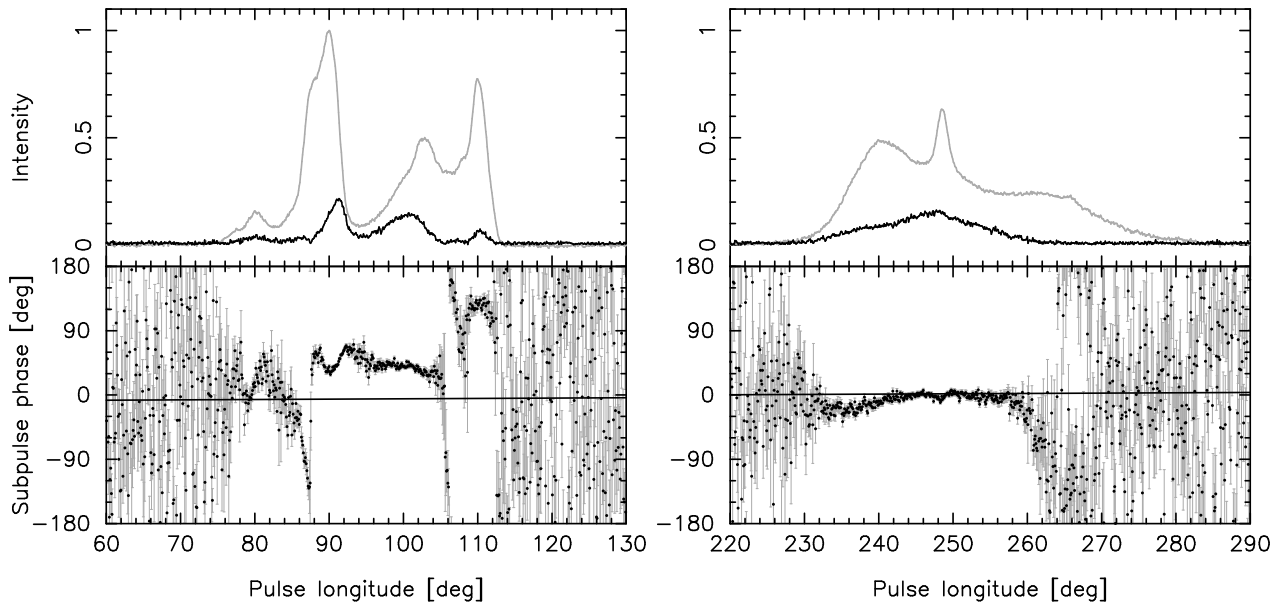}
\caption{\label{Figtscs}The upper panels (MP on the left, IP on the
right) show the strength of the $20P$ modulation pattern as function of
pulse longitude (black curve) and the pulse profile (grey
curve). The lower panels show the corresponding subpulse phase as
function of pulse longitude with an arbitrary phase offset applied to
make the subpulse phase zero at the centre of the IP. The solid line
indicates the slope expected if the pattern is strictly in
phase in the pulsar frame.}
\end{figure*}

The fact that the MP and IP both show a similar periodic modulation
pattern is already a remarkable result in itself. We will explore this
similarity further by determining any phase relations of the $\sim20P$
modulation pattern observed in the different profile components. The
technique used to obtain this phase relation as function of pulse
longitude is based on fluctuation spectra. Firstly the (complex) LRFS
was calculated for 128 pulses long blocks of data. Instead of
calculating the { square} of the complex numbers in the spectrum, as
is done to obtain the LRFS as shown in Fig. \ref{FigModulation}, the
phases were calculated. By selecting the spectral bin that showed the
$\sim20P$ modulation pattern strongest (which was at 0.046875 cpp),
the so-called subpulse phase as function of pulse longitude was
obtained \citep{es02}. The { complex numbers} obtained for the different blocks of
data are added coherently using an iterative scheme designed to
maximize the correlation between the phase profiles of the different
blocks in the on-pulse region. This is achieved by allowing an
arbitrary phase offset, which is constant with pulse longitude,
between the different blocks of data. { The coherent addition gives rise to the complex subpulse modulation envelope, with the complex phase corresponding to the subpulse phase.} Error bars on the subpulse phase were calculated by
applying a bootstrap scheme similar to the one employed in Section
\ref{SectModIndex} for the modulation index.

The result of the calculation is shown in the lower panels of Fig.
\ref{Figtscs}. The actual value of the subpulse phase is not an
intrinsic property of the pulsar, but phase differences between
different pulse longitudes are. We have therefore arbitrarily chosen
the subpulse phase to be zero at the centre of the IP. The upper panels
of Fig. \ref{Figtscs} show, in addition to the pulse profile, the { amplitude}
of the subpulse modulation { corresponding to the amplitude of the complex modulation envelope}\footnote{ The normalisation of the subpulse amplitude profile is such that for a pure sinusoidal signal with a frequency equal to the centre of a frequency bin in the complex LRFS the subpulse amplitude profile matches the pulse profile with its peak amplitude normalised to 1.} (see \citealt{es02,esv03}).  { Note that this amplitude depends on the coherence of the modulation pattern (i.e. the stability of the longitude dependence of the subpulse phase as function of pulse longitude). The results obtained} are
identical in the first and second half of the data and for the data
which was not corrected for the baseline artifact. In addition the
BPSR data gave identical results, although the pulsar was much weaker
during that observation. Hence we are confident that the features of
Fig. \ref{Figtscs}, both at the MP and IP, are stable and intrinsic
properties of the star rather than instrumental effects.

The subpulse phase relation with pulse longitude reveals a high degree
of complexity, especially for the MP. Concentrating on the IP first,
we see that the subpulse phase is roughly independent of pulse
longitude. This means that that the modulation cycle seen in different
pulse longitude bins of the IP are in phase, hence the intensity bands
seen in the pulse-stack are to a good approximation horizontal (at least
on average). This lack of phase drift makes it unclear if the physical
mechanism responsible for the modulation pattern is the same as is
operating in pulsars with drifting subpulses. At the outer edges of
the IP profile the modulation pattern appears to be slightly out of
phase such that the subpulse phase decreases away from the central
parts of the profile. This indicates that the modulation pattern is
slightly lagging at the centre of the IP. The central peak of the IP
at pulse longitude $\sim\degrees{248}$ does reveal a slight deviation
in subpulse phase as well.

In the MP, at the longitudes where the $20P$ modulation is strongest
(pulse longitudes $\degrees{88}-\degrees{105}$), the subpulse phase
appears to be roughly in phase as well, although revealing a weak
decline corresponding to a gradual drift from trailing to leading
components. The subpulse phase offset between the IP and the central
parts of the MP is about $\degrees{40}$, which corresponds to a delay
of $\sim2.5P$ such that the pattern in the MP is late compared to that
of the IP.
It must be noted that if the modulation patterns in two pulse
longitude bins are intrinsically in phase (in the pulsar frame), they
will be observed out of phase because later pulse longitudes are
observed later in time, allowing the pattern to change during the
rotation of the star. The slope in subpulse phase expected to be
observed if the modulation is intrinsically in phase is indicated by
the almost horizontal baseline in Fig. \ref{Figtscs} continued
through from the MP to the IP. We can conclude that the $\sim20P$
modulation cycle of the IP is intrinsically out of phase with the
modulation of the MP at any longitude and that the observed phase
delay of the modulation patterns corresponds to an intrinsic delay of
$\sim3P$. As will be discussed in Section \ref{SectIssues},
this delay corresponds to an extremely long light travel distance
which is much larger than the light-cylinder radius, thereby
challenging existing theories for pulsar emission to explain this.

\begin{figure}
\includegraphics[height=0.95\hsize,angle=270]{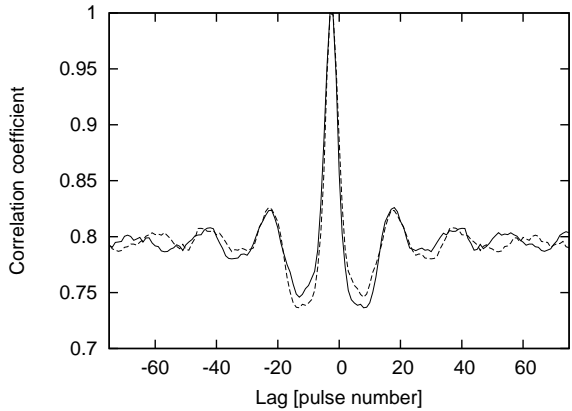}
\caption{\label{Figcorrel}The cross correlation between the pulse
energies of parts of the MP and IP where the $\sim20P$ modulation
pattern is strongest. The peak of the correlation function (solid
curve), which is normalised to 1, peaks at a negative lag. The dashed
curve is the same function, but time reversed with a shift of
$5P$. This formally confirms the phase-locked $\sim2.5P$ lag by which
the MP follows the IP independent of any jitter in the $20P$
modulation.}
\end{figure}

The phase delay between the MP and the IP can be demonstrated in a
different way by correlating the measured pulse energies for
successive pulses in the MP and IP. The cross correlation function,
using the regions in the MP and IP where the modulation is strongest,
is shown in Fig. \ref{Figcorrel}. The $\sim20P$ modulation cycle is
clearly visible in the correlation function, showing that the
modulation patterns in the MP and IP are phase-locked independent of
any jitter in the periodicity of the pattern. The phase offset in the
patterns gives rise to an offset of the peak of the correlation
function from zero lag. The shift can be determined by comparing the
offset between the cross correlation curve and the time reversed
correlation function, which is set to $5P$ in the figure. This offset
is equal to twice the offset of the symmetric curve from zero lag and
therefore confirms the lag of $\sim2.5P$ of the pattern observed in
the MP compared to that in the IP.

The behaviour described so far is accompanied by a number of strong
diversions. The modulation pattern in the trailing peak of the MP (at
pulse longitude $\degrees{110}$) appears to be out of phase with that
in the more central regions by $\sim\degrees{80}$, corresponding to a
delay of almost $5P$. More subtle is the change in subpulse phase in
component II at the location of the strongest presence of the
$\sim20P$ modulation (pulse longitude $\sim\degrees{90}$).  Although
the subpulse phase is approximately longitude stationary in that pulse
longitude range, it decreases towards pulse longitude
$\degrees{90}$, indicating that the pattern is observed early at the
pulse longitudes where the modulation is strongest.  Close examination
of the MP profile in Fig. \ref{Figtscs} reveals that component II is
in fact composed of two components which overlap.
The trailing half of component II shows the $20P$ modulation
strongest, suggesting that only one of the two overlapping components
strongly participates in the modulation cycle.

At the pulse longitudes where the { amplitude of} the $\sim20P$ modulation
pattern is weakest the subpulse phase has the most complex pulse
longitude dependence. In the first half of component II (pulse
longitude $\sim\degrees{86}$) a rapid swing in subpulse phase occurs
which cover almost a complete modulation cycle of $\sim20P$
(corresponding to \degrees{360} in subpulse phase).  Similarly between
components III and IV (around pulse longitude $\degrees{107}$) the
modulation power disappears accompanied by a steep negative slope in
the subpulse phase gradient. We note, however, that the power affected
by this feature is very weak compared with the total power in the
modulation pattern.

The main conclusion of the subpulse phase analysis is that the regions
where the $\sim20P$ modulation pattern is strongest a firm
phase-locking is present between the MP and IP. We note that the delay
accompanied by this phase-locking is found to be identical in
completely different observations recorded at different dates. This
implies that the locking is sustained over periods much greater than
the duration of a single observation and is likely to be a permanent
feature of the pulsar.

\section{Discussion}

\label{SectIssues}
\subsection{The emission of PSR B1055--52}

\subsubsection{Modulation features}

PSR B1055--52 is a pulsar with emission of great complexity. It is therefore important to assess the dominant features of the emission and their implications for possible physical models.

Its most obvious feature is the $\sim{20P}$ modulation at both poles (Fig.~\ref{FigModulation}). Here we have demonstrated that this is not a precise periodicity, nor even a jitter about a specific periodicity, but loosely-connected quasi-periodic sequences. The sequences show no sign of `drifting subpulses', and their intensity bands vary in intensity. In these respects PSR B1055--52 resembles PSR B1702--19 \citep{wws07}, also a near perpendicular rotator. So it may be argued that the lack of drift may be a consequence of being highly inclined.  But regular modulations without drift are also found in PSR B0656+16 (\citealt{wws+06}), which is inclined at merely $\sim\degrees{30}$ \citep{ew01}. So the lack of drift may be simply a consequence of being relatively young (as many highly inclined pulsars are  as according to \citealt{wj08a}).  

The second remarkable feature is the phase-locking of the IP and MP. Inspection of Fig.~\ref{FigSinglePulses} shows that at times much (but not all) of the MP is imitating the IP in detail of intensity and pattern. The intrinsic delay of $3P$ is sufficiently large to be puzzling when measured in terms of light-travel distance (can the signal between the poles have travelled 19 light-cylinder radii without losing coherence?). Yet the delay is small enough compared with the $20P$ modulation that we must conclude that information is conveyed from the IP to the MP, rather than the reverse. 
This would make the IP the `driving' agent { (as originally surmised by \citealt{big90a})}, and cross-correlations between its components show that despite its complex profile the IP effectively operates as a single component. This is illustrated by the subpulse phase plot of Fig.~\ref{Figtscs}, where the IP modulation remains at a single phase, spread across all components. However, in the geometry established in Paper I, the IP is located on open polar field lines (as expected for pulsars' radio emission), and it is not clear how it can `drive' the MP, { which, under the assumption of a simple dipolar structure, is observed on putative} closed field lines well within the light cylinder. 

The full picture represented by Fig.~\ref{Figtscs} is complex and potentially revealing. It is possible that the open field lines on the pole $\emph{opposite}$ to that of the IP may share the IP's modulation but these are anyway invisible to our sightline. 
Note that the MP, for reasons unknown, has a `missing' leading half as shown in Fig. 7 of Paper I. 
This implies that the closest we get to viewing the open field lines { associated with the MP pole} are the field lines on which the MP's leading components (I and II) lie, and with increasing pulse longitude our sightline moves progressively deeper into the `auroral' { zone. Therefore} component IV seems to define a boundary with a much-reduced corotating zone (see Fig. 8 of  Paper I){, although we should note that the implied presence of current flow within such a zone would require a modification of the assumed dipolar field lines}. The $\sim{20P}$ modulation of the MP can be argued to appear at three different phases at three different longitudes.  In components I and the first part of component II  the $20P$ modulation appears weakly and intermittently at zero lag { with respect to that in the IP}, close to the (hidden) open field lines of the second pole and therefore suggestive of synchronised two-pole modulation. Moving deeper into the magnetosphere, and partly overlapping the earlier components, the second half of component II and component III are dominated by the principal modulation at a lag of $2.5P$ (which corresponds to an intrinsic delay of $3P$). A third phase of the $20P$ modulation appears in component IV, where the emission is actually almost $\emph{anti}$-correlated with the IP emission and appears in Fig.~\ref{Figtscs} to $\emph{precede}$ exact anti-correlation by $\sim\degrees{40}$.
This effect can be seen, for example, around pulse 320 in the pulse-stacks of Fig.~\ref{FigSinglePulses}. 
Note that at two points in the profile (one just \degrees{2} before the pulse longitude of peak emission and  one at pulse longitude $\sim\degrees{106}$) 
transitions from correlated to anti-correlated phases are marked by a rapid swing in subpulse phase.
In any case, it is clear that the entire magnetosphere (with the possible exception of the small corotating region) is participating in the $20P$ modulation at multiple phases.

It should be noted that this picture is not necessarily inconsistent with the two-pole model for PSR B1702--19 (\citealt{wws07}). In that pulsar it was argued that modulated emission was intrinsically in phase at both poles and the emission was seen on open field lines from both poles. In PSR B1055--52 open field lines are only visible to us at the IP so that it is quite possible that the pole opposing the IP is in exact phase with the IP. At the MP of PSR B1055--52 our sightline then catches a delayed feedback on field lines which are invisible to us in PSR B1702--19.

There is also an interesting parallel with the behaviour of PSR B1822--09. This pulsar is also most probably a near perpendicular rotator \citep{bmr10} and has two distinct modes of emission (`B' and `Q'; \citealt{fwm81}).  In the B-mode the IP disappears and the MP gains an additional component, so these components are anti-correlated \citep{fw82,gjk+94}.  Even within the Q-mode the IP is anti-correlated with one of the MP subcomponents (see Fig. 11 of \citealt{bmr10}). This is analogous to the behaviour of the IP and component IV in PSR B1055--52 and may have a common physical explanation. 

\subsubsection{Pulse characteristics}

{  Figures \ref{FigSinglePulses} and \ref{FigEnDist} give us very different views of the emission. In the former we see a contrast between the spiky MP and broader IP emission, while in the latter we note a near-identical MP and IP energy distribution. The energy distribution's lognormal structures (a common feature of pulsars; \cite{bjb+12}) imply a random multiplicative process, and the fact that they have the same standard deviation appears to be telling us that identical processes are at work throughout the magnetosphere. 

However, we found no correlation between the MP and IP fluxes at timescales longer than the modulation cycle, so the fact that the distributions have the same mean intensity (corresponding to the same MP and IP flux in the integrated profile) must be attributed to chance, and cannot favour our sightline and observing frequency and cannot be the result of reflections or refractions of emission from a single source. On the other hand, in the pulse-stacks of figure \ref{FigSinglePulses} we see the consequence of our particular sightline. So the appearance of a pulse, spiky or smeared, could depend on how our sightline samples the pulsars field lines, while the form of the energy distributions are related to the underlying processes.

We also note that PSR B1055--52 is not the only pulsar which shows radio pulses with different characteristics in their MP and IP. \cite{he07} have shown that the giant pulses observed in the IP of the Crab pulsar are much broader compared to those in the MP, reminiscent of the broader subpulses seen in the IP of PSR B1055--52. However, the IP pulses observed in the Crab pulsar show a very different observing frequency structure compared to those in the MP, while our data shows no such effect. Hence it is unclear if these two pulsars are linked or not.
}

\subsection{Seeking a physical model}
\label{SectPhysicalModel}

The essentially geometric results of Paper I, and those presented hitherto in this paper, make no physical assumptions other than that we are observing a rotating dipole emitting polarized radiation parallel to its field lines.  To develop a physical understanding, we need to explain (1) the origin of the dominant $\sim20P$ modulation cycle, (2) why there is a considerable  phase-locked delay between the IP emission and that at the MP and (3) why this pulsar emits radiation on field lines which are normally deemed to be closed. 
\subsubsection{Intrinsic models}
The polar cap model (\citealp{rs75,gs00}), based on a discharging vacuum gap above the poles, would have difficulty in accounting for the diverse emission of PSR B1055--52. There is no evidence either in the integrated profile or the single pulse behaviour to suggest an underlying conal structure and associated drifting subpulses such as is found in so many, generally older, pulsars \citep{ran86,wes06,wse07}. However, it has been argued \citep{gs00} that in relatively young pulsars the polar sparks will be narrow and spiky and display intensity modulation rather than drift. { Spiky emission} is indeed found at the MP of PSR B1055--52 (though not at the IP). Nevertheless, all versions the polar cap model require the presence of non-dipolar field line structures close to the star's surface, possibly with a local sunspot-like structure, to give sufficient field line curvature to trigger pair-creation \citep{gs00}.  However, this argument can be turned on its head: why would two very different local poles then have magnetic structures which generate precisely the same $P_3$ periodicities? Furthermore, the poles share the same value of $P_3$, even though its precise value varies intermittently around $20P$. This, and the exact synchronised phase-lock delay of $2.5P$, demands some form of inter-pole communication which the polar cap model does not describe in its present form. 

By contrast, a model of non-radial neutron star oscillations (adapted by \citealp{cr04} from the asteroseismology of magnetic white dwarfs) provides a natural way of inter-pole phase-locking via the surface of the neutron star. This has already been pointed out in the similar case of PSR B1702--19 \citep{wws07}. In this model, the neutron surface is seen as having multiple narrow concentric strips centred on the magnetic poles which oscillate at a rapid frequency, with each at 180$\deg$ phase shift from its neighbour. The $20P$ emission modulation then results from beats between the oscillation frequency and the observer's sampling rate of  $1/P$, and could be expected to be the same at both poles.  The phase shift of the modulation is harder to explain: this should either be zero (as it is in the case of PSR B1702--19) or 180$\deg$, at least in the simplest version of the model.  Neither a zero or \degrees{180} phase shift is consistent with the intrinsic $3P$ delay observed in PSR B1055--52. However, an advantage of this model is that it can plausibly allow for oscillations beyond the confines of the polar cap. The polar cap boundary is conventionally defined by the size of the light cylinder, and not by the neutron star's surface properties which would determine the oscillations. Thus emission in the closed magnetosphere is theoretically possible -- but it would still be necessary to explain why PSR B1055--52 is apparently so exceptional in this respect. Note that \cite{kjw+09} have argued for emission outside the conventional open field line region for other interpulse pulsars.

If we insist that information between the poles is magnetosphere-borne, the intrinsic delay time of $\sim{3P}$ is extraordinarily long. In his original study of this pulsar, \citealp{big90a} pointed out that the light travel distance along a field line from one pole to the other is unlikely to be more than $\sim{1.1 P}$. An intrinsic delay time of $\sim{3P}$ corresponds to a light travel distance of about 19 light-cylinder radii. One way to achieve this would be for the polar emission to undergo reflections within the magnetosphere and/or some kind of magnetosphere-wide drift before becoming observable at the opposite pole \citep{wri03,my11}. \cite{wri03} invokes interactions between the pulsar's polar region and the null surfaces. In a highly-inclined pulsar the null surface can stretch to great heights above the pole along the rotation axis \citep{rw10} and drifting particles may encounter electric gaps some 10 light-cylinder from the poles. Pair-generated particles would take a round trip time of the order of the intrinsic $3P$ delay time observed in PSR B1055--52 and may well return to regions well away from the polar gaps. In a more general paper \cite{my11} argue strongly that corotating `dead' zones cannot exist in an inclined rotating pulsar magnetosphere due to the innate inability of the pulsar to screen the inductive electric field. Hence particle drift must occur throughout the magnetosphere and not only on open field lines. Although these authors only detail their model for nearly-aligned pulsars it has obvious wider application to PSR B1055--52 and its ideas would be well worth pursuing.      

\subsubsection{Extrinsic models}
It seems that any model based on an isolated neutron star has difficulty in replicating the full complexity of the observations. Could it be that the complexity arises from the intrusion of extraneous material into the magnetosphere? \cite{lm07} have outlined such a model, in which neutral material leaks into the corotation region (\citealp{tsy77,wri79}) and, following ionisation, modulates the radio emission through (ir)regular instabilities. Auroral currents would cause the field lines to drift compared with corotation and thereby generate periodic modulations and quite plausibly generate a profile of multiple components. 
However, one would expect both poles to be modulated at almost the same phase (i.e much less than a single period), and, again, the rigid phase difference of $2.5P$ between poles remains hard to explain. 
 
A debris disk akin to the intrusion model has been suggested by \cite{cs08} as an explanation for intermittent modulations of pulsar radio emission. Such a `fallback' disk \citep{md81} is argued to arise as neutral supernova material which has remained gravitationally bound to the neutron star gradually spirals inwards, spinning ever more rapidly as the star is approached, finally crossing the light cylinder. Individual rocks of various sizes are held together through their tensile strength and thus survive tidal disruption, despite gradual erosion through surface ionisation. For our present purposes this model has the advantage of introducing a natural source of periodic modulations at roughly the timescales required. In PSR B1055--52 at just under half a light cylinder radius the orbital period of such material would be around 20 times the stars rotation period. Thus well within the light cylinder, yet outside the deduced closed region and perhaps sufficiently far from the neutron star to escape destruction by evaporation, we could expect the range of periods we observe as pulse modulation. The disk could comprise several distinct rings of fragmented neutral material (weak modulation), and/or be dominated by a small number of larger asteroids (strong modulation).  An important advantage of this model is that an orbiting body might be viewed with an orbital time-delay at opposing poles and thereby explain the phase-locked delay. Furthermore, as might be expected from a passing source, the $\sim{20P}$ modulation appears in the pulse-stack (Fig. \ref{FigSinglePulses}) as short-lived streaks of emission. 

The attractive features of the disk model made us further investigate its consequences. One critical expectation of the model is that the $\sim{20P}$ modulations should ultimately reflect an underlying rigid orbital periodicity, however much random jitter becomes imposed on it by the pulsar's magnetosphere. We therefore took the brightest sequence of our data (the pulse-stack shown in Fig. \ref{FigSinglePulses}) and tested it systematically with folds of the IP emission at different periodicities in the range $18-20P$. 
As discussed in Section \ref{SectStability} the possibility of a fixed orbital timescale is ruled out.

\section{Summary and Conclusions}
\label{SectSummary}

In Paper I we were able to establish that PSR B1055--52, with its unusual and highly asymmetric integrated radio profile (Fig.~\ref{Fig1055profiles20}), belongs to the relatively small class of pulsars where we are viewing radio emission on field lines emanating from both magnetic poles. Such pulsars present us with a rare opportunity to test emission theories, especially if, as here, their single-pulse emission exhibits specific modulation patterns at each pole. 

The radio emission of PSR B1055--52 defies any traditional pulsar category: there is little support for the picture of symmetrically-placed nested emission cones and an associated pattern of subpulse drift. Instead we find a phase-locked delay between the modulations at opposite poles -- and periodic modulation occurring on field lines normally attributed to a corotating `dead' zone. Furthermore, the emission of virtually every component of both the MP and IP has a different characteristic, ranging from smooth periodic to `spiky' unmodulated.

These results represent a major conundrum for pulsar theorists.  The $\sim{20P}$ modulation found at both poles, although not precisely periodic, is phase-locked over the many epochs spanning our observations and therefore almost certainly a permanent lock. In the profile components at both poles where this modulation is strongest, the pattern takes the form of one or two strong pulses followed by little emission until the next band. In this respect, the emission resembles that of PSR B1702--19 \citep{wws07}, a pulsar similar in age and inclination -- but whose interpulse peak separation is close to \degrees{180} and we can be confident we are observing emission on open field lines above the polar caps at alternate poles. By contrast, at one of the poles of PSR B1055--52 (the MP) we are observing emission on field lines which are well away from the polar cap and deep within what is normally deemed to be the closed co-rotating zone.

The most striking feature of our present study is the magnitude of the phase-locked delay between the IP and MP modulations. At $\sim{3}P$ it implies relativistic communication distances far exceeding the pulsar's light cylinder radius (corresponding to a light travel time of $0.16P$), yet is short compared with the  $\sim{20P}$ modulation it delivers. In the case of PSR B1702--19 the intrinsic phase lock was  virtually simultaneous
and it was possible to argue \citep{wws07} that the poles may communicate to one another through the body of the neutron star itself (possibly via non-radial oscillations).  Now this argument is much harder to make.

Differences in the chance orientation of the observer's geometry may be responsible for the different appearances of PSRs B1702--19 and B1055--52. It is possible that both pulsars emit their quasi-periodic signals simultaneously at both poles on open field lines but are only observable in PSR B1702--19. These then later (in both pulsars) generate a response signal from certain closed field lines but are only observable in PSR B1055--52. The simple uniform-phase nature of the IP's modulation compared to the complex multi-phase response of the MP (Fig. \ref{Figtscs}) combined with the fact that the emission pattern of the MP is lagging that of the IP strongly suggests that the IP somehow $\emph{drives}$ the MP.  We argue from our observations that the $20P$ bursts are magnetosphere-wide phenomena,  synchronised at both poles and simultaneously affecting regions well away from the poles. 

As an explanation for the phase-locking we initially found the idea of a circumstellar disk very appealing. An `asteroid' at a radius well within the light-cylinder can indeed have an orbital period commensurate with the observed $20P$ modulation, and this would be a natural way to bring about a phase-lock of significant delay. However, attempts to find the underlying orbital period by folding the modulation sequences met with failure: the $\sim{20P}$ modulations are not a simple jitter superimposed on precise periodicity.

The popular (and in many cases convincing) polar cap model \citep{rs75,gs00} has difficulties in explaining inter-pole locking. The model requires the emission modulation patterns to be fixed by local conditions at each magnetic pole and does not involve magnetosphere-wide interaction.  Even if polar cap sparks can account for the emission behaviour of older `drifting' pulsars, in the cases of PSRs B1055--52 and B1702--19 something quite different is needed. Possibly a model akin to those of \citep{wri03,my11} could work, linking separated regions of the magnetosphere and causing the entire magnetosphere to participate in the subpulse drift. Alternatively, it is conceivable that irregularities in the pulsar's electrical current balance might be responsible.

It should be noted that PSRs B1702--19 and B1055--52 are not the only known examples of inter-pole communication. The IP of PSR B1822--09 was shown to interact with its MP many years ago (\citealp{fw82,gjk+94}) and its geometry has recently been confirmed as a near-perpendicular rotator by \cite{bmr10}. In this pulsar a periodic modulation ($\sim42P$) is again shared by both the MP and IP, and the IP itself only appears during the one of the pulsar's two modes.  There is evidence \citep{lhk+10} that the mode-change is accompanied by a change in the spin-down rate. Clearly, the modulated radio emission of near-perpendicular rotating pulsars, although intrinsically weak, is a diagnostic of powerful events.
 
Taken together, the locked and subtle interactions between the poles of PSRs B1055--52, B1702--19 and B1822--09  must contain major clues to the nature of the elusive pulsar emission mechanism and the structure of the pulsar magnetosphere. We have here focused on the single-pulse behaviour of PSR B1055--52 and briefly considered a range of models. None satisfy all the observational constraints, so much work remains for those seeking to understand pulsars.

\section*{Acknowledgments}
{ We would like to thank Ben Stappers as well as the referee of this manuscript for useful discussions.}
GW thanks the University of Sussex for the ongoing award of a Visiting Research Fellowship and is grateful to the CSIRO for support during a stay at ATNF. The Australia Telescope is funded by the Commonwealth of Australia for operation as a National Facility managed by the CSIRO. 

\bibliographystyle{mn2e}

\label{lastpage} 

\clearpage

\end{document}